\documentclass[12pt,preprint]{aastex}
\def\fuse{{\it FUSE\/}}                 
\def\stis{STIS}
\def\hst{{\it HST\/}}
\def\lya{Ly${\alpha}$}

\def\EE#1{\times 10^{\small#1}}

\def\teff{$T_{\rm{eff}}$}
\def\logg{log~$g$}
\def\iue{{\it IUE\/}}
\def\euve{{\it EUVE\/}}
\def\nhi{$N$(H\,I)\/}
\def\lnhi{log~$N$(H\,I)\/}
\def\lnoi{log~$N$(O\,I)\/}
\def\noi{$N$(O\,I)\/}

\def\kms{km s$^{-1}$}
\def\arcs{\ifmmode {''}\else $''$\fi}

\def\mfarcs{\hbox{$~\!\!^{\prime\prime}$}}

\shorttitle{O/H in the Local Bubble}
\shortauthors{Oliveira et al.}

\received{2002 October 31}
\begin{document}

\title{O/H in the Local Bubble\altaffilmark{1}}
%{Based on observations made with the NASA-CNES-CSA Far Ultraviolet Spectroscopic Explorer. {\it FUSE} is operated for NASA by the Johns Hopkins University under NASA contract NAS5-32985.}}

\author{Cristina M. Oliveira\altaffilmark{2}, Jean Dupuis\altaffilmark{2}, Pierre Chayer\altaffilmark{2}\altaffilmark{,3}, and H.~Warren~Moos\altaffilmark{2}}
  
\altaffiltext{1}{Based on observations made with the NASA-CNES-CSA {\it Far Ultraviolet Spectroscopic Explorer}. \fuse~is operated for NASA by The Johns Hopkins University under NASA contract NAS5-32985. Based also on observations made with the NASA/ESA Hubble Space Telescope, obtained from the Data Archive at the Space Telescope Science Institute, which is operated by the Association of Universities for Research in Astronomy, Inc., under NASA contract NAS 5-26555.}
\altaffiltext{2}{Department of Physics and Astronomy, The Johns Hopkins University, Baltimore, MD 21218}
%\altaffiltext{3}{Institut d\'{}Astrophysique de Paris, 98$^{bis}$ boulevard Arago, F-75014 Paris, France}
%\altaffiltext{4}{Current Address: Center for Astrophysics and Space Sciences, University of California at San Diego, C-0424, La Jolla, CA, 92093}
\altaffiltext{3}{Primary affiliation: Department of Physics and Astronomy, University of Victoria, P.O. Box 3055, Victoria, BC V8W 3P6, Canada}

\begin{abstract} % OK OK

We present new measurements of the oxygen gas-phase abundance along four sightlines probing gas inside the Local Bubble, using data obtained with the $Far~Ultraviolet~Spectroscopic~Explorer$ ($FUSE$)~and the $Hubble~Space~Telecope$ ($HST$). Combining these with seven previously published $N$(O\,I) and $N$(H\,I) measurements we determine a mean O/H ratio for the Local Bubble, (O/H)$_{\rm LB}$ = (3.45 $\pm$ 0.19)$\times10^{-4}$ (1$\sigma$ in the mean). Our result is virtually identical to O/H = (3.43 $\pm$ 0.15)$\times10^{-4}$ derived from data for sightlines probing gas well outside the Local Bubble. In contrast to the D/O and D/H ratios, which seem to have different values beyond the Local Bubble, our results show that the O/H ratio is constant and single-valued both inside and outside the Local Bubble, for low density sightlines, with column densities up to log $N$(H) $\simeq$ 21. In addition, the similarity of the two values above suggests that the net effect of the uncertain O\,I $f$-values in the $FUSE$~bandpass is not significant. Taking into account the latest determinations of the Solar abundance of oxygen, our result implies that $\sim$25\% of the oxygen in the Local Bubble is depleted onto dust grains. The similarity of the value for O/H along low average density sightlines in the Local Bubble with that of denser sightlines beyond may permit a limit on the diluting effects of infalling low metallicity gas.

\end{abstract}


\section{INTRODUCTION} % OK OK

%Oxygen, the third most abundant element in the interstellar medium (ISM) after hydrogen and helium, plays a vital role in our understanding of galactic chemical evolution, interstellar chemistry and dust grains formation and composition. 
The abundance of oxygen in the interstellar medium (ISM) is an important parameter of galactic chemical evolution models. In addition, oxygen, the third most abundant element in the ISM after hydrogen and helium, plays an important role in interstellar chemistry and in dust formation models.

\citet{1998ApJ...493..222M} used the weak intersystem $\lambda$1356 O\,I transition to determine O/H along the line of sight to 13 stars (with distances ranging from 130 to 1500 pc), with log $N$(H\,I) = 20.18 -- 21.28 and log $f$(H$_{\rm 2}$) = $-$5.21 -- $-$0.20. They found a remarkably constant O/H = (3.43 $\pm$ 0.15)$\EE{-4}$ (where the $f$ value has been adjusted; see \S\ref{disc}). \citet{2003ApJ...591.1000A} used the same O\,I transition to determine O/H toward 19 stars (with distances ranging from 0.84 to 5.01 Kpc) with log $f$(H$_{\rm 2}$) = $-$1.52 to $-$0.33 and log $N$(H\,I) = 20.94 -- 21.41, deriving O/H = (4.08 $\pm$ 0.13)$\times10^{-4}$ (1$\sigma$ in the mean).
More recently, \citet{2004ApJ...613.1037C} used the same transition to determine O/H toward 36 new sightlines, with distances ranging from $\sim$0.5 to 6.5 Kpc. Combining the new results with previously published $HST$ results they find a bimodal distribution in O/H, with O/H = (3.90 $\pm$ 0.11)$\EE{-4}$ along low density sightlines ($n_{\rm H} \leq$ 1 cm$^{-3}$) and O/H = (2.84 $\pm$ 0.12)$\EE{-4}$ along denser sightlines. In addition, these authors find, for the low density sightlines ($n_{\rm H} \leq$ 1 cm$^{-3}$), that the O/H ratios diverge when one considers pathlengths shorter or longer than 800 pc. They find (O/H)$_{\rm short}$ = (3.47 $\pm$ 0.07)$\times10^{-4}$, vs, (O/H)$_{\rm long}$ = (4.37 $\pm$ 0.09)$\times10^{-4}$. 
  
In the Local Bubble, $N$(O\,I) is low enough that the $\lambda$1356 transition is undetected; its $f$-value is $\sim$13 times weaker than that of the weakest, measurable, O\,I transition in the $FUSE$~bandpass ($\lambda$974). For sightlines within the Local Bubble $N$(O\,I) can only be determined with $FUSE$~data. The $FUSE$~bandpass contains numerous O\,I transitions, with different oscillator strengths, which allow accurate O\,I column densities to be determined. The strong $\lambda$1302 transition ($f$ = 4.80$\times10^{-2}$), available only with STIS, is always saturated, and as such, oxygen column densities derived solely from this transition are not reliable.

The studies by \citet{1998ApJ...493..222M} and \citet{2003ApJ...591.1000A} indicated that O\,I is a good tracer of H\,I in the nearby Galactic disk. However, only a few sightlines inside the Local Bubble have published O/H ratios \citep[see][and references therein]{2002ApJS..140....3M,2003ApJ...587..235O}. 

%This study was motivated by the fact that O seems to be a good tracer of H in diffuse sightlines and using D/O to estimated D/H for LB with O/H from Meyer leads to low D/H when we know from direct measurements that D/H is higher.
%There are only a few measurements of O/H for the local bubble so we carried out a systematic study of O/H in the local bubble using data from...

In this paper, we use data obtained with $FUSE$~and the Space Telescope Imaging Spectrograph (STIS), onboard $HST$, in conjunction with previously published O\,I and H\,I column densities, to determine the gas-phase O/H ratio in the Local Bubble, (O/H)$_{\rm LB}$. The observations and data reduction are described in \S\ref{obs}. Determinations of $N$(O\,I) and $N$(H\,I) are presented in \S\ref{noimeasurement} and \S\ref{nhimeasurement}, respectively. The results are presented in \S\ref{resultsanddiscussion} and discussed in \S\ref{disc}. Our finding are summarized in \S\ref{sum}. All results are quoted at the 1$\sigma$ level, unless otherwise noted.

\section{OBSERVATIONS AND DATA PROCESSING} % OK OK
\label{obs}

We use data obtained with \fuse~to determine \noi~and data obtained with \stis~onboard \hst~to determine the neutral hydrogen column densities. The stellar properties of the 4 stars are summarized in Table \ref{star_properties}. The observations are described with details below.

\subsection{\fuse~Observations}

The \fuse~observatory consists of four co-aligned prime-focus telescopes and Rowland-circle spectrographs that produce spectra over the wavelength range 905 -- 1187~\AA~with a spectral resolution of $\sim$~15~--~20~km~s$^{-1}$ (wavelength dependent), for point sources. Two of the optical channels employ SiC coatings, providing reflectivity in the wavelength range $\sim$~905~--~1000~\AA,~while the other two have LiF coatings for maximum sensitivity above 1000~\AA. Dispersed light is focused onto two photon-counting microchannel plate detectors. % With this arrangement of optical channels (LiF 1, LiF 2, SiC 1, and SiC 2) and detector segments (1A, 1B, 2A, 2B) the \fuse~instrument has 8 segments: LiF 1A, LiF 1B, LiF 2A, LiF 2B, SiC 1A, SiC 1B, SiC 2A, and SiC 2B. Four channels cover the wavelength range 1000~--~1080~\AA~while two channels each cover the ranges 900~--~1000~\AA~ and 1080~--~1180~\AA. 
 The \fuse~mission, its planning, and on-orbit performance are discussed by \citet{2000ApJ...538L...1M} and \citet{2000ApJ...538L...7S}

Table \ref{fuse_obs} summarizes the \fuse~observations of the four targets studied in this work. The data were obtained through both the large (LWRS, $30\mfarcs\times30\mfarcs$) and medium sized apertures (MDRS, $4\mfarcs\times20\mfarcs$) in histogram and time-tagged modes (HIST and TTAG, respectively in Table \ref{fuse_obs}). The two-dimensional \fuse~spectra are reduced using the CalFUSE pipeline v2.4.1\footnote{The CalFUSE pipeline reference guide is available at http://fuse.pha.jhu.edu/analysis/pipeline\_reference.html}. The processing includes data screening for low quality or unreliable data, thermal drift correction, geometric distortion correction, heliocentric velocity correction, dead time correction, wavelength calibration, detection and removal of event bursts, background subtraction, and astigmatism correction.
The spectra are aligned by cross-correlating the individual exposures over a short wavelength range that contains prominent spectral features and then coadded by weighting each exposure by its exposure time, using the CORRCAL software developed by S. Friedman. All the spectra are binned to three pixel samples, or $\sim$20 m\AA, for analysis (the line spread function, LSF, is about 11 pixels or $\sim$70 m\AA~wide). For each target, observations obtained through the same aperture are coadded in order to increase the S/N of the dataset.

\subsection{STIS Observations}
\label{stis_calib}

Table \ref{stis_obs} summarizes the STIS data for the stars. %Note that $N$(H\,I) for the WD\,0032+035 sightline has been determined by \citet{2000ApJ...544..423V}. We will be using their value in our analysis (see discussion {\bf below}).
The data are reduced using the standard STSDAS pipeline within IRAF. The CALSTIS pipeline \citep[version 2.3][]{stispipeline} is used to reduce the data obtained with the E140H and E140M echelles. 
%The CALSTIS pipeline (version 2.3) is used to produce the two-dimensional spectrum of the FUV-MAMA and then the scattered-light removal is performed using different algorithms for the different echelle gratings. For the E140M data the \citet{2000AAS...197.1202L} algorithm is used to estimate and remove the scattered light from the data in the pipeline procedure. 
The echelle spectral orders in the regions adjacent to \lya~are combined using a weighted averaging scheme where the orders overlap.
% For the E140H data the scattered light is removed with the procedure of \citet{2000AJ....119.2481H}.
For stars where there are multiple observations the different exposures are combined by weighting them by the exposure time, to increase the S/N of the data. We use a single-Gaussian with a FWHM of 1.2 and 1.3 pix to describe the line spread function of the E140H and E140M data, respectively.

\section{Determination of $N$(O\,I)}
\label{noimeasurement}

We use apparent optical depth, curve of growth and profile fitting methods (hereafter AOD, COG, and PF, respectively) to derive $N$(O\,I). We follow the procedures outlined in \citet{2003ApJ...587..235O} to use these techniques, with the following modification: a single-Gaussian with a FWHM of 10.5 pixels (constant across all wavelengths and channels and during the fitting procedure) is used to describe the \fuse~line spread function. Table \ref{eqwidth} summarizes the wavelengths and $f$-values \citep{2003ApJS..149..205M} of the O\,I transitions used in this study, as well as their measured equivalent widths. Table \ref{methods} presents, for each star, $N$(O\,I) derived with the different methods, as well as the adopted $N$(O\,I) values. We indicate in parentheses which line was used to determine the AOD result which is quoted in Table \ref{methods}. For the AOD measurements we use the weakest O\,I line detected along each sightline, when the derived column densities have reasonable uncertainties, such as for WD\,0050$-$332 and WD\,1254$+$223. For WD\,0232$+$035 and WD\,2152$-$548, the AOD results quoted are from stronger lines. For these sightlines, AOD measurements on the weakest lines are consistent with those of the stronger lines used, which provide smaller uncertainties in the column densities. The adopted $N$(O\,I) values represent compromises between the O\,I column densities obtained with the different methods. We chose to, subjectively, adopt $N$(O\,I) values that are the midpoint of the range of measurements obtained with the different methods, with uncertainties that overlap with the extreme measurements. We note that no molecular hydrogen was detected along any of the four sightlines presented here. Below we comment briefly on each line of sight independently.

%\subsection{Apparent Optical Depth}

%To yield accurate column densities the AOD method must be used on weak lines, i.e., lines which are in the linear part of the COG. By comparing column densities derived from two transitions which differ in strength ($f\times\lambda$) by a factor close to two (preferably) one is able to determine if there is unresolved saturation in these lines \citep[see][for a thorough discussion of this method]{1991ApJ...379..245S}. Unlike the COG and PF methods, weak lines are not sensitive to unresolved structure along the line of sight. The weak lines require a good S/N to yield reasonable uncertainties in $N$(O\,I). Due to atomic physics, the weaker O\,I transitions in the \fuse bandpass occur in the SiC channels which have a lower reflectivity than the LiF channels, hence a lower intrinsic S/N. In cases where the COG (or PF) results differ substantially from the AOD results (see discussion below), perhaps due to unresolved structure, we chose to adopt the column densities derived with the AOD method, albeit at the expense of larger uncertainties.

%\subsection{Curve of Growth}

%The curve of growth uses lines on the linear part together with lines on the saturated parts of the COG to determine a best fit to the measured equivalent widths in terms of $N$ and $b$. The stronger lines, which constrain $b$ have typically a lower measurement uncertainty than the lines which fall on the linear COG. When there is unresolved saturated structure

\subsection{WD\,0050$-$332}

The weighted average of the AOD measurements (on the S1B and S2A channels) on $\lambda$924.9 yields \lnoi~=~15.15 $^{+~0.09}_{-~0.12}$, consistent with AOD on $\lambda$929.5 (\lnoi~=~15.18 $^{+~0.07}_{-~0.08}$). The COG fit yields \lnoi~=~14.98 $^{+~0.04}_{-~0.04}$. The STIS E140H data for this sightline shows at least two components in N\,I and Si\,II separated by $\sim$10 km s$^{-1}$. The velocity of the stronger component, $v_{\odot}$ = 6.8 \kms, is consistent with the expected velocity of the LIC along this direction \citep{1995A&A...304..461L}, $v_{\rm LIC}$ = 6.2 $\pm$ 1.2 \kms. Using the double-component velocity structure derived from these species, and fitting N\,I, Si\,II and O\,I together we find log $N$(O\,I) = 15.09 $\pm$ 0.09.  %Even though our COG results agrees with the AOD and PF results within the uncertainties we chose to adopt the AOD results which are less sensitive to the velocity structure along the line of sight and which are independent of the LSF of the instrument used. 
We adopt then  \lnoi~=~15.07 $\pm$ 0.09. We use more data than \citet{2003ApJ...595..858L} which find \lnoi~=~15.26 $^{+~0.06}_{-~0.05}$. 
%We adopt \noi~=~15.15 $\pm~^{0.09}_{0.12}$ for this line of sight, from a combination of single and double component profile fitting, AOD and COG measurements. 
%Even though we have more data than \citet{2003ApJ...595..858L} we derive a consistent \noi. 15.28+0.08-0.07 here versus 15.26+0.06-0.04.  AOD (924.9) yields 15.04 +0.14-0.20 S2A, 15.29+0.11-0.14 S1B, with a weighted average of 15.15+0.09+0.12. AOD on 929.5 yields 15.17+0.11-0.09 S1B and 15.19+0.10-0.12 S2A, with a weighted average of 15.18+0.07-0.08.
%BUT in the STIS data there are at least two comps separated by 10km/s which are visible in Si II and NI. Using this v structure and fitting together NI and Si II with OI I get N(OI) = 15.09 +/- 0.09 which yields a much normal O/H than before (for the OI lines the LSF=10.5 pix fixed). The derived T and vturb for the two comps is very reasonable. The 1 comp fit (fix LSF=10.5pix) on 1039, 976, 948, 936, 929, 924.9 yields 15.05+0.06-0.02. 
%PF of 976, 948, 936, 929, 924 with T=10K fix yields:
%LSF=3.0 N=15.12, b=15.0, LSF=3.5 N=15.13 b=13.7; LSF=4.0 N=15.14 b=11.6; LSF=4.5 N=15.19 b=7.4, LSF=4.75 N=15.28 b=6.3, LSF=5.0 N=15.29 b=6.3. 
%f*lambda ratio for 929.5/924.9 is 1.48 so assume 924.9 is not saturated and adopt the AOD weighted avg. result for this star: 15.15+0.09-0.12.
%If I wanted to use weighted avg of AOD on 924.9 and 929.5 this would be 15.17+/-0.06.
%DONE

\subsection{WD\,0232$+$035}

\begin{figure}
\begin{center}
\epsscale{0.85}
%\rotatebox{90}{
\plotone{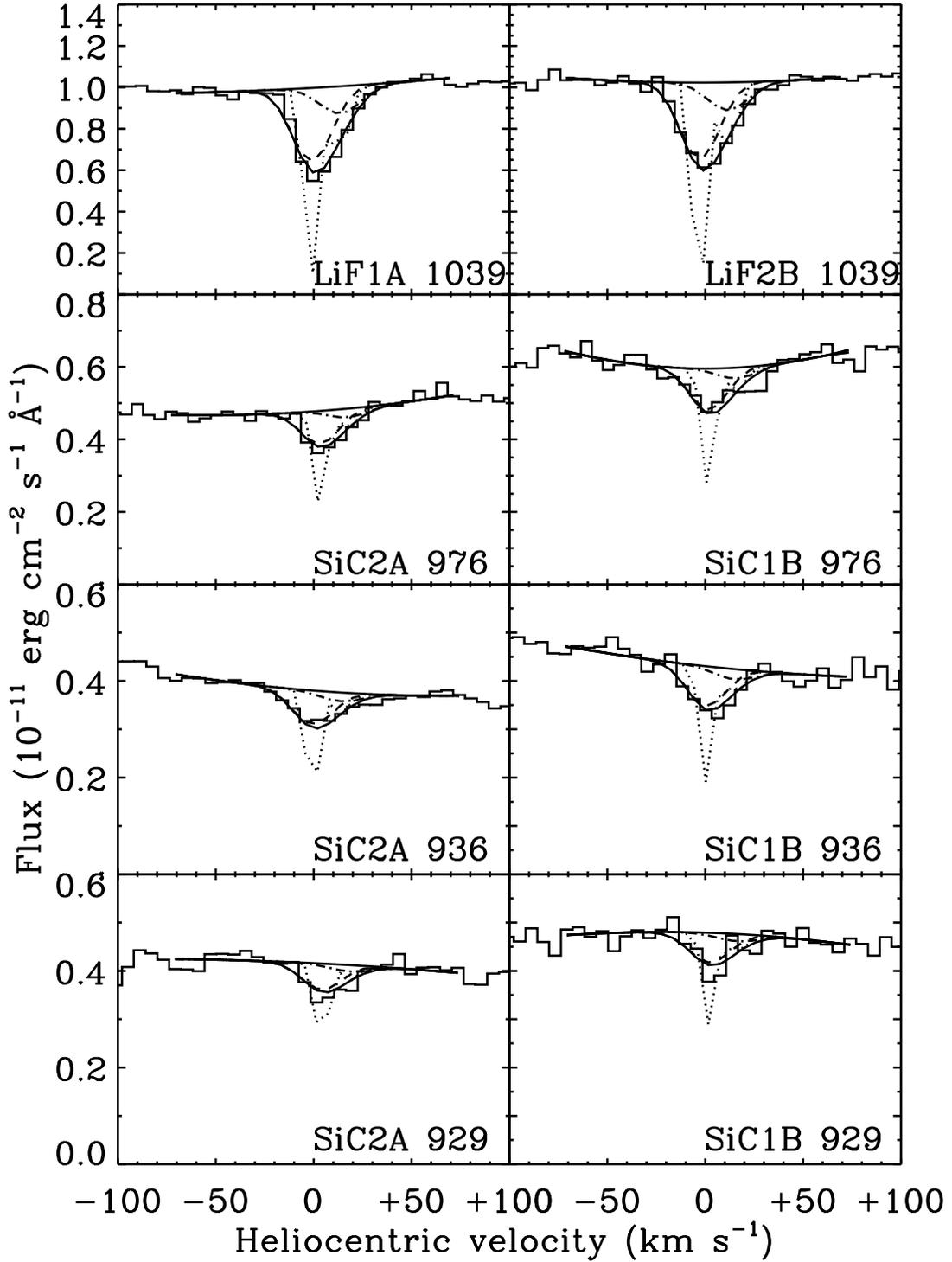}
%}
\caption{Fits to some of the O\,I lines used in PF for the WD\,0232$+$035 sightline, with the two-component model discussed in the text. Dotted lines represent the fit prior to convolution with the instrument LSF, dashed lines and dot-dash lines represent the two components. Solid lines represent the adopted continua as well as the fit after convolution with the instrument LSF. \label{gd153lines}}
\end{center}
\end{figure}

AOD measurements on the $\lambda$976 O\,I transition yields \lnoi~=~15.01 $\pm$ 0.03. 
The COG fit to O\,I transitions only in the \fuse~bandpass yields \lnoi~=~15.06 $^{+~0.07}_{-~0.06}$. 
% with $b$ = 5.7 $\pm~^{0.8}_{0.6}$ \kms. 
When the $\lambda$1302 O\,I transition (present in the STIS data) is included in the COG fit we find \lnoi~=~14.96 $^{+~0.04}_{-~0.03}$. \citet{2000ApJ...544..423V} have analyzed the STIS data for this star. They find a stronger component at $v_{\odot}~+$3.1 \kms, and a weaker component at $v_{\odot}~+$17.6 \kms. The latter component is identified with the LIC. We use the two-component velocity structure derived by these authors to fit the O\,I transitions in the \fuse~bandpass. Using PF we derive \lnoi~=~14.94 $^{+~0.05}_{-~0.04}$. Figure \ref{gd153lines} presents fits to some of the O\,I lines used with profile fitting, with the two-component model discussed above. Dotted lines represent the fit prior to convolution with the instrument LSF, and dashed lines indicate the adopted continua. We adopt then \lnoi~=~15.00 $\pm$ 0.06 for this sightline, as a compromise between the AOD, COG and PF measurements.

\subsection{WD\,1254$+$223}
%{\bf needs work}

The low column density along this line of sight, combined with the S/N of the data allows us to detect only the $\lambda$988 (triplet) and $\lambda$1039 O\,I transitions in the \fuse~data. Since the $\lambda$988 transition is strongly contaminated with terrestrial airglow, we use only the weak $\lambda$1039 transition in conjunction with the strong $\lambda$1302 line (from the STIS data) to constrain $N$(O\,I) along this line of sight.

We find \lnoi~=~14.22 $\pm$ 0.03, from a weighted average of the AOD measurements (on four channels) on the $\lambda$1039 transition. The COG fit to the $\lambda$1039 and $\lambda$1302 transitions yields \lnoi~=~14.27 $\pm$ 0.04. We adopt then \lnoi~=~14.25 $^{+~0.06}_{-~0.05}$. Our result is consistent, within the uncertainties, with \lnoi~=~14.38 $^{+~0.20}_{-~0.12}$ from \citet{2003ApJ...595..858L}. Their result is based on the possible detection of the $\lambda$976 transition which is $\sim$3 times weaker than $\lambda$1039 ($\tau_{0,\rm1039}$ = 0.24), not detected in our dataset. The difference between the two datasets might be due to the CalFUSE pipeline used; \citet{2003ApJ...595..858L} used CalFUSE v2.0.5, while we use v2.4.1.

%But NL says he also sees 976 (which I don't!) and that AOD on this line leads to 14.38+0.20-0.12, suggesting saturation of 1039. (fl)976=3.23, (fl)1039=9.56; 1039 is 2.96 stronger than 976. The peak optical depth in 1039 L1A is 0.24, so this does seem to indicate that 1039 is not saturated

%To justify that 1039 doesn't not suffer from airglow contamination I looked at the L1A data extracted through the MDRS (this is a LWRS obs) for each observation. There is no hint of airglow cont. in the L1A MDRS. In addition comparison of the L1A 1039 profile for each individual coadded observation indicates that the profiles are similar, so if there were some airlgow you would expect the profiles to be different since the airglow should not be constant between observations.
%DONE

\subsection{WD\,2152$-$548}

We derive \lnoi~=~15.35 $^{+~0.08}_{-~0.10}$ using the AOD method on the $\lambda$936 transition. The COG and PF techniques yield \lnoi~=~15.52 $^{+~0.07}_{-~0.06}$ and \lnoi~=~15.30 $^{+~0.13}_{-~0.09}$, respectively. We adopt \lnoi~=~15.45 $\pm$ 0.10 in good agreement with \lnoi~=~15.49 $^{+~0.10}_{-~0.07}$ derived by \citet{2003ApJ...595..858L}.

\section{Determination of $N$(H\,I)}
\label{nhimeasurement}

We use observations of the H\,I \lya~transition obtained by the Space Telescope Imaging Spectrograph onboard $HST$~in conjunction with stellar models, to determine $N$(H\,I) along three of the four sightlines studied in this work. Below we discuss each star separately. The adopted $N$(H\,I) are presented in Table \ref{ratios}.

\subsection{WD\,0050$-$332}

\subsubsection{Stellar Model}

%To measure the interstellar H\,I column density, the stellar continuum in the neighborhood of \lya~has to be considered. Th

Several authors have used the Balmer line series to determine \teff~and \logg~for this white dwarf. Using pure-hydrogen LTE stellar models \citet{1997ApJ...488..375F}  find \teff~=~35,816 $\pm$ 148 K and \logg~=~7.679 $\pm$ 0.031, while \citet{1997ApJ...480..714V} find \teff~=~36,320 $\pm$ 200 K and \logg~=~7.92 $\pm$ 0.03. \citet{2003MNRAS.341..870B} have also analyzed the H\,I Balmer lines, using non-LTE pure-hydrogen stellar models. These authors find \teff~=~35,660 $\pm$ 135 K and \logg~=~7.93 $\pm$ 0.03. To take into account the stellar H\,I absorption a set of nine pure-H NLTE stellar models corresponding to \teff~=~(35,000;36,000;37,000) K and \logg~=~(7.65;7.80;7.95) was constructed. The range in the stellar parameters, \teff~and \logg, was chosen to include the published \teff~and \logg~values in the literature, as well as their associated uncertainties. 

We use the velocity shift between photospheric and interstellar absorption determined by \citet{2003MNRAS.341..477B} to align the stellar model with the data. From an analysis of STIS E140M and \iue~data, these authors find $v_{\rm PH} - v_{\rm ISM} \sim +$25 km s$^{-1}$, corresponding to a shift of $\sim$18 pix. 
 
%{\bf DID NOT USE CHRIS SUBTRACTION PROGRAM BECAUSE DIDN'T LIKE WHAT IT DID TO LyA}

\subsubsection{Ly$_{\alpha}$~Profile Fitting}

\begin{figure}
\begin{center}
\epsscale{0.50}
\rotatebox{90}{
\plotone{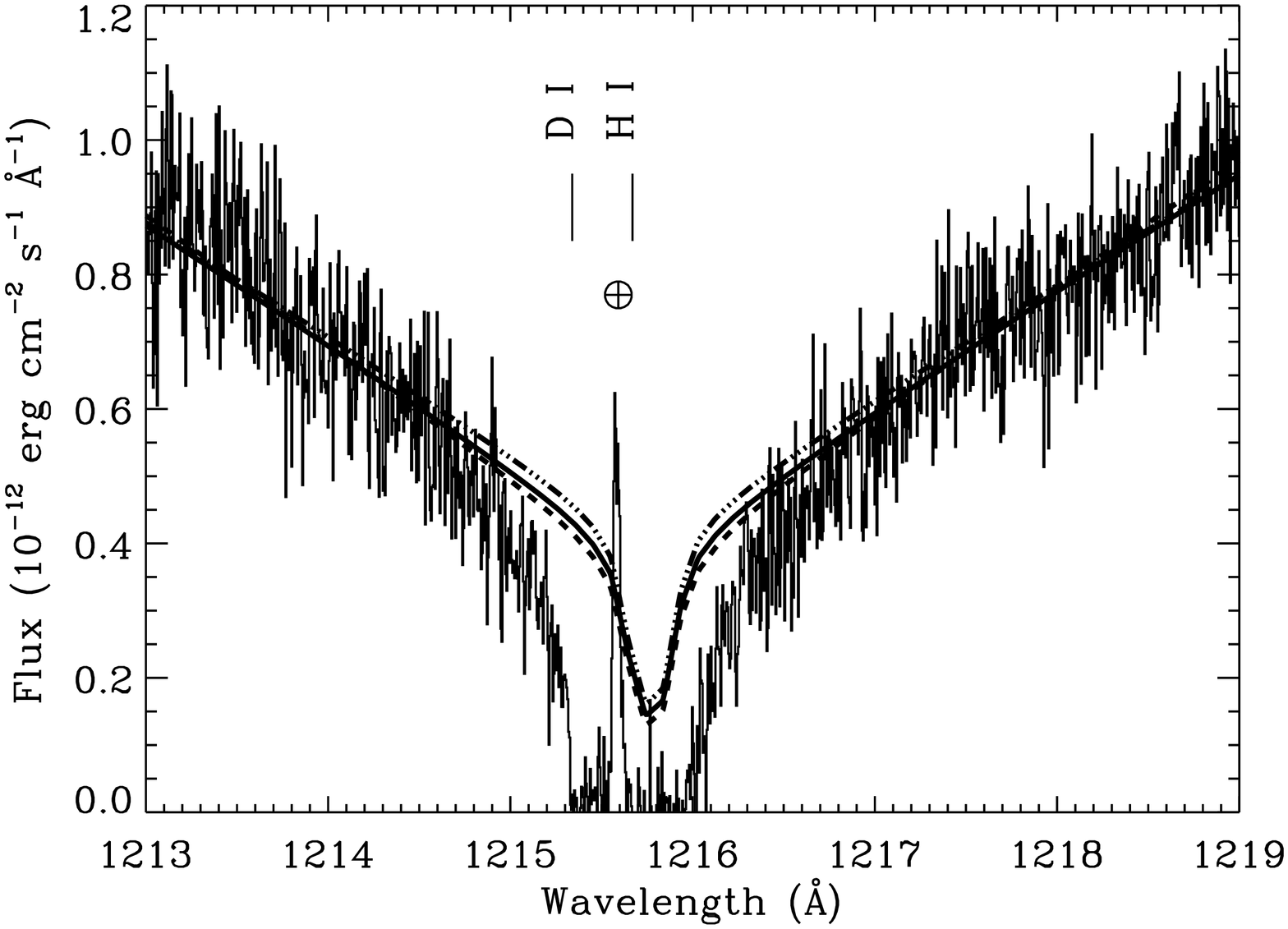}
}
\caption{H\,I \lya~absorption for WD\,0050$-$332. The histogram shows the coadded and merged STIS E140H observations of this line of sight; the thick solid line shows the adopted stellar model, corresponding to \teff~=~36,000 K and \logg~=~7.80. Also displayed are the two models which produce the lowest \nhi~(\teff~=~35,000 K, \logg~=~7.65, dashed line) and the highest \nhi~(\teff~=~37,000 K, \logg~=~7.95, dashed-dotted line). Geocoronal emission is annotated with $\oplus$.\label{wd0050-332_lya}}
\end{center}
\end{figure}

We follow the procedure outlined in \S\ref{stis_calib} to calibrate and coadd the STIS data covering the \lya~region. Profile fitting of the H\,I \lya~transition alone without constraining the Doppler parameter, $b$, leads to unreasonable estimates of $N$ and $b$. We use then the Lyman edge, covered by \fuse~in the SiC1B channel, to constrain $b$ which is then used as a constraint when fitting the \lya~region. By fitting the H\,I Lyman series transitions between 913.0 \AA~and 919.2 \AA~with the continuum fixed, we determine $b_{\rm H\,I}$ = 8.0 $^{+~0.1}_{-~0.2}$ km s$^{-1}$. In all our fits we use the fitting code {\bf Owens} \citep[for a more detailed description of this code see][]{2002ApJS..140..103H,2002ApJS..140...67L}. After aligning and normalizing the data by each of the stellar models described above fit the \lya~region. No D\,I is included in the fit and the $b$-value is fixed at 8.0 \kms. In addition, a first-order polynomial is used to model the continuum after normalization by the model atmosphere, to take into account uncertainties in the instrument sensitivity. Figure \ref{wd0050-332_lya} shows the best fit model, corresponding to \teff~=~36,000 K and \logg~=~7.80, overplotted on the coadded data. We show also the two models which produce the most extreme H\,I column densities, which lead to our 1$\sigma$ systematic uncertainties. These are the models with \teff~=~35,000 K, \logg~=~7.65 and \teff~=~37,000K, \logg~=~7.95. We find \lnhi = 18.57 $\pm$ 0.02 $^{+~0.01}_{-~0.02}$, where the first uncertainties quoted take into account the statistical uncertainties associated with using the best fit model (\teff~=~36,000 K, \logg~=~7.80) and the second quoted uncertainties reflect the systematic uncertainties associated with using different stellar models. Using a conservative approach to combine the two uncertainties we adopt \lnhi = 18.57 $^{+~0.03}_{-~0.04}$. The value of $b_{\rm H\,I}$ derived above, by fitting the H\,I Lyman series transitions, implies $T_{\rm max}$ = 3840 K, considerably lower than the mean temperature, $T$~= 6680 $\pm$ 1490 K, derived by \citet{2004ApJ...613.1004R} from an analysis of sightlines within 100 pc. To ascertain the influence of $b_{\rm H\,I}$ on $N$(H\,I) we performed multiple fits using the best fit stellar model and fixing $b_{\rm H\,I}$ to different values. We find that $N$(H\,I) agrees at the 1$\sigma$ level with the value quoted above, for $b_{\rm H\,I} <$ 20 km s$^{-1}$. Considering typical $b$-values for H\,I along similar sightlines \citep{2004ApJ...613.1004R}, 20 km s$^{-1}$ is a reasonable upper limit for $b_{\rm H\,I}$. Hence our derived $N$(H\,I) should not be biased by the $b_{\rm H\,I}$ adopted.

Our adopted \nhi~is in good agreement with the values determined by \citet{1997MNRAS.286...58B} and \citet{1999A&A...346..969W} (\lnhi~=~18.49 $^{+~0.10}_{-~0.12}$ and \lnhi~=~18.51 $^{+~0.07}_{-~0.13}$, respectively) from analyses of \euve~data for this star.

%For this line of sight there is also a degeneracy between \nhi and $b$ value. I found \nhi = 18.40 $\pm$ 0.03 and $b$ = 25.8 $\pm~^{0.6}_{0.7}$ km s$^{-1}$ with $b$ free.
%This $b$ value is then used as a constraint in determining $H$(H\,I) using the \lya~transition.

\subsection{WD\,1254$+$223}

\subsubsection{Stellar Model}

Determinations of the effective temperature and gravity for this white dwarf range from \teff~=~38,686 K, \logg~=~7.662 \citep{1997ApJ...488..375F} to \teff~=~41,200 K, \logg~=~7.85 \citep{1992ApJ...390..590V}. 

To determine the influence of the different stellar models in the measured interstellar H\,I column density we constructed a grid of nine pure-H NLTE stellar models corresponding to the values \teff~= (39,000; 40,000; 41,000) K and log $g$ = (7.60; 7.75; 7.90). \citet{2003MNRAS.341..477B} tentatively determined a velocity shift of $v_{\rm PH} - v_{\rm ISM}$~=~$+$21 km s$^{-1}$ between the photospheric and the interstellar absorption. We use the shift determined by these authors to align the stellar models with the data. 

%This corresponds to a shift of $\sim$16 pix
%We coadded the different exposures obtained from the mast archive, did not use Chris program to remove background. DID NOT LOOK WHAT CHRIS PROGRAM DID .  

\subsubsection{\lya~Profile Fitting}

\begin{figure}
\begin{center}
\epsscale{0.50}
\rotatebox{90}{
\plotone{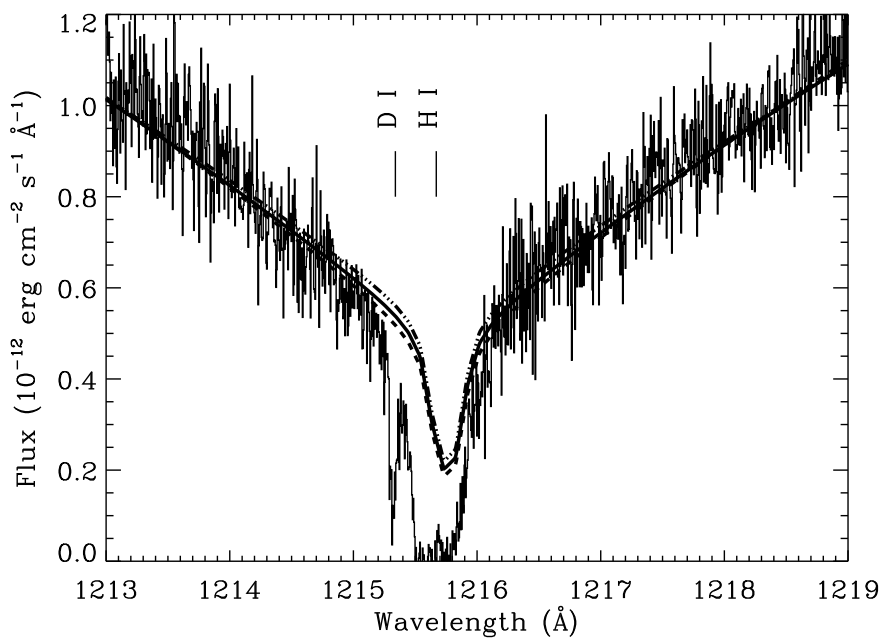}
}
\caption{\lya~absorption for WD\,1254$+$223. The histogram shows the coadded and merged STIS E140H observations of this line of sight; the thick solid line shows the best fit stellar model (\teff~=~40,000 K and log $g$ = 7.75). The other two models plotted produce the lowest (\teff~=~39,000 K, \logg~=~7.60, dashed line) and highest (\teff~=~41,000 K, \logg~=~7.90, dashed-dotted line) \nhi. \label{gd153_lya}}
\end{center}
\end{figure}

For each of the nine computed stellar models we normalize the data by the stellar model prior to profile fitting. The D\,I \lya~transition is also clearly visible in the data (see Figure \ref{gd153_lya}) so we fit both D\,I and H\,I. Because H\,I can sometimes trace gas not detected in other species we fit the \lya~D\,I and H\,I profiles in separate absorption components. We use a second-order polynomial to model the continuum after normalization by the model atmosphere. The Doppler parameter is a free parameter of the fit and we derive $b_{\rm H\,I}$ = 11.6 $^{+~0.7}_{-~0.5}$ km s$^{-1}$ and $b_{\rm D\,I}$ = 7.6 $^{+~1.1}_{-~0.7}$ km s$^{-1}$. The $b$-values derived here for H\,I and D\,I are consistent with the temperature and turbulent velocity ($T$~=~7000 $^{+~2900}_{-~2800}$ K, $\xi$~=~1.2 $^{+~1.1}_{-~1.2}$ km s$^{-1}$) derived for this line of sight by \citet{2004ApJ...613.1004R}. 
%, consistent with the Doppler parameter being dominated by thermal rather than turbulent broadening. 
Figure \ref{gd153_lya} displays the data overplotted with the best fit model (\teff~=~40,000 K and \logg~=~7.75) and the two models that produce the lowest \nhi~(\teff~=~39,000 K and \logg~=~7.60) and the highest \nhi~(\teff~=~41,000 K and \logg~=~7.90). We find log$N$(H\,I) = 17.85 $\pm$ 0.03 $\pm$ 0.02, where the first uncertainties are statistical and the second are due to the different stellar models. We use a conservative approach to combine the two uncertainties and adopt log$N$(H\,I) = 17.85 $\pm$ 0.05. Our measured \nhi~is in good agreement with previous determinations using \lya~data obtained with the \iue~satellite \citep[\lnhi~=~17.92 $\pm$ 0.10,][]{1998ApJS..119..207H} and \euve~data \citep[][\lnhi~=~17.90, \lnhi~=~17.89 $\pm$ 0.04, and \lnhi~=~17.90 $\pm$ 0.04, respectively]{1996aeu..conf..217D,1997MNRAS.286...58B,1999A&A...346..969W}. %Although it is beyond the scope of this paper to do a full analysis of $N$(D\,I) along this line of sight, $N$(D\,I) derived in our analysis leads to a D/H ratio that is consistent with the Local Bubble value quoted by \citet{2004ApJ...609..838W}. This gives us more confidence in our $N$(H\,I) determination.

%\clearpage

\subsection{WD\,2152$-$548}

\subsubsection{Stellar Model}

\begin{figure}
\begin{center}
\epsscale{0.50}
\rotatebox{90}{
\plotone{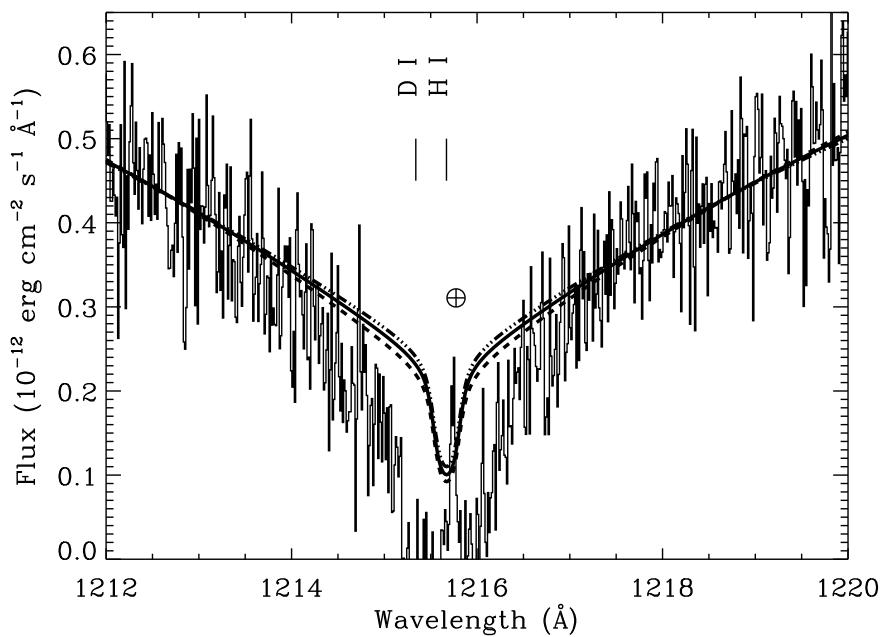}
}
\caption{Interstellar H\,I \lya~toward WD\,2152$-$548. The histogram shows the STIS E140M observations of this line of sight; the thick solid line shows the adopted stellar model, corresponding to $T_{\rm eff}$ = 46,000 K and log $g$ = 7.85. Also displayed are the models that produce the lowest (\teff~=~44,000 K, log $g$ = 7.70, dashed line) and highest (\teff~=~48,000 K, log $g$ = 8.00, dashed-dotted line) $N$(H\,I). Geocoronal emission is annotated with $\oplus$.\label{wd2152_lya}}
\end{center}
\end{figure}

Determinations of \teff~for this white dwarf range from \teff~=~44,000 K \citep{1997ApJ...488..375F} to \teff~=~46,000 K \citep{1997ApJ...480..714V} while \logg~ranges from 7.8 to 7.9 \citep{1997ApJ...480..714V,1997ApJ...488..375F}. To take into account these different determinations of \teff~and \logg~as well as their associated uncertainties we constructed a grid of nine pure-hydrogen NLTE stellar models corresponding to \teff = (44,000;46,000;48,000) K and \logg = (7.70;7.85;8.00). Figure \ref{wd2152_lya} displays the stellar model with \teff~=~46,000 K and \logg~=~7.85 overplotted on the STIS E140M data for the \lya~region.  
%The data was divided by the stellar model prior to profile fitting the \lya~region. The derived \nhi

\subsubsection{\lya~Profile Fitting}

For each of the nine stellar models computed we normalize the data by the stellar model prior to profile fitting the \lya~H\,I transition (no D\,I is included in the profile fit). We use a single-Gaussian LSF with a FWHM of 1.3 pixels. In addition, a second-order polynomial is used to model the continuum after normalization by the model atmosphere, to take into account uncertainties in the instrument sensitivity. \citet{2003MNRAS.341..477B} determined that $v_{\rm PH} - v_{\rm ISM} \sim -$12 km s$^{-1}$. Profile fits taking into account the velocity shift between photospheric and interstellar absorptions have shown that this shift is negligible and thus does not affect the interstellar \nhi~determination.

As with WD\,0050$-$332, profile fitting of the \lya~transition alone without constraining the $b$ value leads to unreasonable estimates of $N$ and $b$. To determine the $b$ value we use the \fuse~data from the SiC 1B channel which covers the Lyman edge. By fitting the H\,I Lyman series transitions between 913.6 \AA~and 915.5 \AA~with the continuum fixed, we determine $b_{\rm H\,I}$ = 8.5 $^{+~2.1}_{-~0.4}$ km s$^{-1}$. This $b$ value is then used as a constraint in determining \nhi~using the \lya~transition. We find \lnhi = 18.84 $^{+~0.05}_{-~0.04}$ $\pm$ 0.01. The uncertainties quoted are statistical (associated with using the best fit model, \teff~=~46,000 K; \logg~=~7.85), and systematic, respectively. Using a conservative approach to combine the two uncertainties we adopt \lnhi = 18.84 $^{+~0.06}_{-~0.05}$. Similarly to WD\,0050$-$332 our derived $b_{\rm H\,I}$ implies $T_{\rm max}$ is in disagreement with the mean temperature derived by \citet{2004ApJ...613.1004R}. To determine the influence of $b_{\rm H\,I}$ in the derived $N$(H\,I) we performed multiple fits using the best fit stellar model and fixing $b_{\rm H\,I}$ at different values. We find that $N$(H\,I) agrees with our adopted value at the 1$\sigma$ level for $b_{\rm H\,I} <$ 25 km s$^{-1}$. Hence our derived \lnhi = 18.84 $^{+~0.06}_{-~0.05}$ is not sensitive to the assumed $b_{\rm H\,I}$, for reasonable values of $b_{\rm H\,I}$.

 Our \nhi~determined from \lya~is in excellent agreement with that determined by \citet{1999A&A...346..969W} from \euve~data, who find \lnhi~=~18.84 $^{+~0.12}_{-~0.16}$, using pure-hydrogen LTE models with \teff~=~44,300 K and \logg~=~7.91.

\section{The Abundance of Oxygen in the Local Bubble}
\label{resultsanddiscussion}

The first part of Table \ref{ratios} presents the column densities and ratios derived in this work. In the second part of Table \ref{ratios} we present a compilation of $N$(O\,I) and $N$(H\,I) (and their respective ratios) from the literature, for sightlines probing only the Local Bubble. Since the ionization balance for both O and H is linked by resonant charge exchange reactions \citep[see][and references therein]{2000ApJ...538L..81J} and there is no H$_2$ along these sightlines, O\,I/H\,I should be representative of O/H along these lines of sight. Hence, in Table \ref{ratios}, O\,I/H\,I has been replaced by O/H. The last column in the table provides information about the sources of the column densities. The choice of the stars listed in the second part of Table \ref{ratios} is based on the contours of the Local Bubble which have been mapped by \citet{1999A&A...346..785S} and \citet{2003A&A...411..447L} using absorption-line studies of Na\,I. They found a cavity with a radius between 65 and 250 pc, that is delineated by a sharp gradient in the neutral gas column density with increasing radius, a dense neutral gas ``wall''. \citet{1999A&A...346..785S} quote an equivalent width of 20 m\AA~of Na\,I $\lambda$5891.59 as corresponding to log $N$(H\,I) $\sim$ 19.3. This $N$(H\,I) is used as a rough estimate of the Local Bubble boundary throughout this work and as such we only searched the literature for stars with log $N$(H\,I) $\leq$ 19.3. We note that in the literature there are other stars that satisfy this criterion \citep[e.g.,][]{2004ApJ...602..776R}. However, for these stars, only the strongly saturated $\lambda$1302 O\,I transition was used to determine $N$(O\,I). We suspect that those column densities are not as reliable as the ones derived here (see note in Table \ref{ratios}) and as such we chose not to include them in Table \ref{ratios}. H{\' e}brard et al. (in preparation) have also demonstrated, based on the constancy of the D/O ratio in the Local Bubble \citep{2003ApJ...599..297H}, that $N$(O\,I) derived by \citet{2004ApJ...602..776R} are unreliable.

\citet{2002A&A...394..691W} have also derived O/H along the sightline to WD\,1029$+$537, which probes only gas in the Local Bubble. We chose not to include their value in Table \ref{ratios} because we believe that their $N$(O\,I) determination might be based on saturated lines ($\lambda$1039 and $\lambda$988.8/988.6). In addition, the inconsistency between the derived $b$-values for the O\,I transitions used in their analysis, $b_{\rm O\,I}$~=~11.2 $\pm$ 1 km s$^{-1}$ ($\lambda$1039) and $b_{\rm O\,I}$~=~30 $\pm$ 5 km s$^{-1}$ ($\lambda$988.8/988.6), leads us to believe that $N$(O\,I) is not reliable. 

All stars listed in Table \ref{ratios} are white dwarfs, but Capella and $\alpha$ Vir.  White dwarfs are particularly useful for studying the ISM in the Local Bubble, because they provide a source of light that covers a wide spectral window ranging from the extreme ultraviolet to the optical. The relatively simple spectra of white dwarfs are also another asset when analyzing absorption lines formed in the ISM. The intense gravitational field that characterizes white dwarfs leads to the downward diffusion of elements heavier than the dominant constituent of the atmosphere (either H or He), leaving the atmosphere devoid of heavy elements.  Therefore, a typical spectrum of white dwarf shows a smooth continuum with only broad hydrogen or helium lines. The simplicity of white dwarf atmospheres minimizes the problem of blend between ISM and photospheric lines.  Observations show, however, that traces of metals are detected in some white dwarf atmospheres. The presence of these heavy elements is attributed to radiative forces or weak stellar winds that counteract efficiently the gravitational settling. For stars showing the presence of heavy elements, a comparison between a stellar model and the observations allows us to identify ISM lines that are blended with photospheric lines.  Consequently, no O\,I lines selected for determining the O\,I column densities presented here are blended with photospheric lines.

The H\,I column densities along the lines of sight toward the white dwarfs listed in Table \ref{ratios} are obtained by fitting the damping wings of the interstellar H\,I Ly$\alpha$ profile. The only exception is WD\,1634$-$573 for which $N$(H\,I) has been obtained by fitting the $Extreme~Ultraviolet$ $Explorer$ ($EUVE$) spectrum with a stellar model and by adjusting the H\,I, He\,I and He\,II column densities \citep[see][and references therein]{2002ApJS..140...91W}. Figures \ref{wd0050-332_lya}, \ref{gd153_lya}, and \ref{wd2152_lya} illustrate that the damping wings of the ISM H\,I Ly$\alpha$ profile span a small portion of the photospheric absorption, hence accurate stellar atmosphere models are needed to reproduce the stellar contribution to the Ly$\alpha$ profile, in order to obtain the ISM profile. This is accomplished by computing stellar atmosphere models under the assumption of non-LTE and by incorporating detailed photospheric Ly$\alpha$ profiles. We note that \citet{2002ApJ...571L.169V} have suggested that for lines of sight with $N$(H\,I) $<$ 10$^{19}$ cm$^{-2}$, H\,I column densities based on the Ly$\alpha$ transition might be overestimated by as much as 20\%, due to the presence of hot H\,I components. However, \citet{2000ApJ...528..756L} compared $N$(H\,I) values derived from $HST$ and $EUVE$ data and found them to be in good agreement.
% This would imply O/H $\approx$ 4.14$\times10^{-4}$.  

The four ratios derived here are in good agreement with previously determined Local Bubble ratios, presented in the second part of Table \ref{ratios}. The O\,I transitions used to derive $N$(O\,I) have been used in other works \citep[see for example][]{2003ApJ...599..297H}. In addition tests have shown that even though some O\,I $f$-values might be slightly inaccurate, the errors should decrease in the final result when multiple O\,I transitions are used \citep{2002ApJS..140..103H}. H\,I column densities have been determined from the Ly$\alpha$ transition numerous times (see for example the $N$(H\,I) references in the second part of Table \ref{ratios}). The good agreement between our $N$(H\,I) and the ones determined with a completely different method using $EUVE$~data, gives us confidence in our $N$(H\,I) determinations.
 
\begin{figure}
\begin{center}
\epsscale{0.70}
\rotatebox{90}{
\plotone{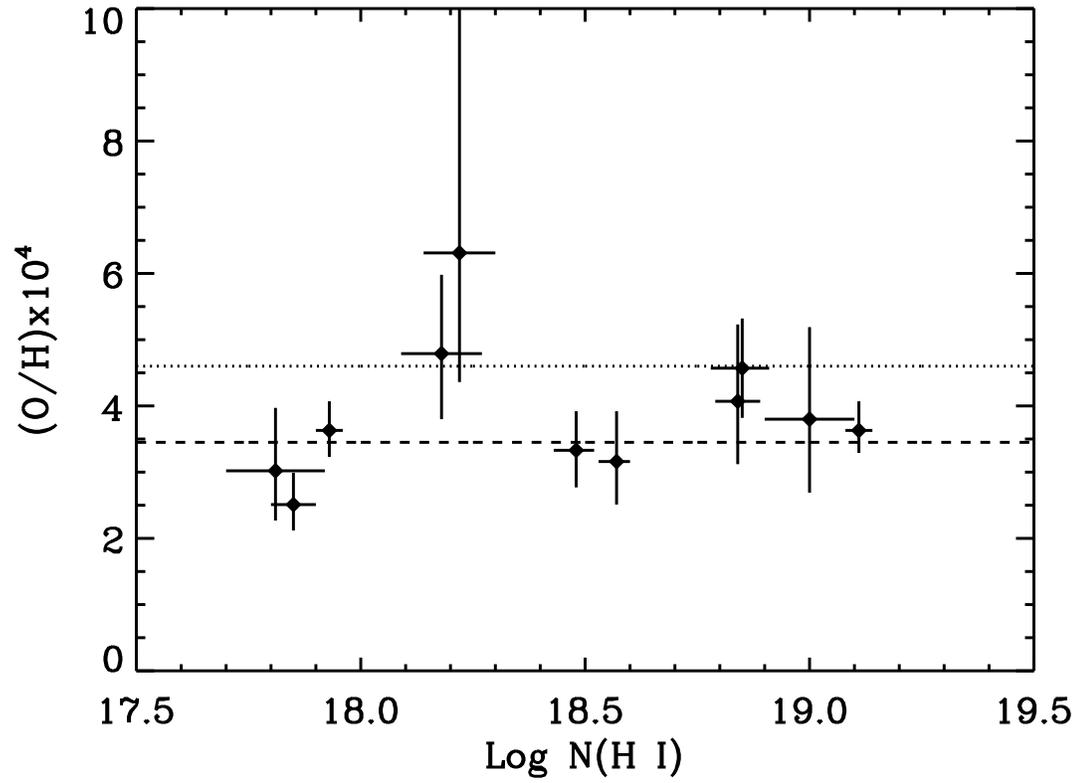}
}
\caption{O/H ratios used in deriving (O/H)$_{\rm LB}$ = (3.45 $\pm$ 0.19)$\times10^{-4}$ (dashed line), as a function of the hydrogen column density. The Solar value (O/H)$_{\odot}$ = 4.6$\times10^{-4}$ from \citet{2004A&A...417..751A} is indicated by a dotted line.\label{ohplot}}
\end{center}
\end{figure}

Using all the ratios in Table \ref{ratios} (except the one for which a lower limit is quoted) we derive the weighted average (O/H)$_{\rm LB}$ = (3.45 $\pm$ 0.19)$\times10^{-4}$ (1$\sigma$ in the mean; average of individual uncertainties used to compute $\sigma$) with $\chi^{2}_{\nu}$ = 1.1 for 10 degrees of freedom. The dispersion around the mean as measured by the square root of the weighted average variance is 0.65$\times10^{-4}$. Figure \ref{ohplot} displays the ratios used to compute the weighted average above, as well as the derived (O/H)$_{\rm LB}$ (dashed line) and the Solar value from \citet{2004A&A...417..751A} (dotted line).

To determine whether any of the ratios dominates the $\chi^{2}$ statistics we determined the $\chi^{2}_{\nu}$ repeatedly, with all the ratios but one. We found that the ratio that most influences the $\chi^{2}_{\nu}$ is the one derived towards WD\,1254$+$223, O/H = (2.51 $^{+~0.48}_{-~0.39}$)$\times10^{-4}$. When this ratio is excluded from the weighted average we find O/H = (3.66 $\pm$ 0.21)$\times10^{-4}$ with $\chi^{2}_{\nu}$ = 0.6 for $\nu$ = 9. The corresponding square root of the weighted average variance is 0.49$\times10^{-4}$. However, this new ratio is still consistent at the 1$\sigma$ level with the weighted average determined with the first eleven ratios in Table \ref{ratios}, O/H = (3.45 $\pm$ 0.19)$\times10^{-4}$. Because the uncertainties vary from sightline to sightline, some of the measurements have a small influence on the value of the weighted mean,  the uncertainty in the mean and the $\chi^{2}_{\nu}$. Thus, the sightlines with the largest uncertainties, G191$-$B2B, $\alpha$ Vir, and Capella, change the weighted mean by only $\sim$2\% and decrease the uncertainty in the mean by $\sim$3\%.

Based on our new and previous measurements of O/H, we conclude that this ratio is constant in the Local Bubble, and that (O/H)$_{\rm LB}$ = (3.45 $\pm$ 0.19)$\times10^{-4}$.

\section{Discussion}
\label{disc}

Our (O/H)$_{\rm LB}$ = (3.45 $\pm$ 0.19)$\times10^{-4}$ is virtually identical to O/H = (3.43 $\pm$ 0.15)$\times10^{-4}$ \citep[][updated $f$-value]{1998ApJ...493..222M} derived using a single transition ($\lambda$1356), for 13 lines of sight probing gas within 1500 pc (most within 500 pc), with 20.18 $\leq$ log $N$(H\,I) $\leq$ 21.28. The $f$-value for $\lambda$1356 ($f$ = 1.16$\times10^{-6}$) as suggested by \citet{1996atpc.book.....W} and \citet{1999ApJS..124..465W} is based on the mean of the lifetimes measured by several workers \citep{1972PhRvA...5.2688J,1974PhRvA...9..568W,1978PhRvA..17.1921N,1990MeScT...1..596M} and the theoretical branching ratio determined by \citet{1992A&A...265..850B}. Our result is also in agreement with O/H = (3.47 $\pm$ 0.07)$\times10^{-4}$ derived by \citet{2004ApJ...613.1037C} for sightlines within 800 pc with average density $n_{\rm H} < 1.0$ cm$^{-3}$.

The $f$-values of the O\,I transitions used to determine all the $N$(O\,I) summarized in Table \ref{ratios} have not been experimentally determined and as such are subject to larger uncertainties. However, the similarity of the O/H ratio determined here with that of \citet{1998ApJ...493..222M} seems to suggest that the net effect of the uncertain $f$-values on $N$(O\,I) is not significant \citep[but see][for two $f$-values, not used here, that are clearly overestimated]{2003ApJ...599..297H}.

By comparing our gas-phase ISM oxygen abundance with the Solar value, and by assuming that the Solar value represents the total (gas + dust) abundance of oxygen, we can estimate the amount of oxygen that can be incorporated into dust.
The commonly adopted Solar oxygen abundance compiled by \citet{1989GeCoA..53..197A}, (O/H)$_{\odot} \sim$8.5$\times10^{-4}$, has been revised downwards during the last few years. \citet{2001sgc..conf...23H} determined (O/H)$_{\odot} \sim$5.5$\times10^{-4}$ and \citet{2001ApJ...556L..63A} (O/H)$_{\odot} \sim$4.9$\times10^{-4}$. More recently \citet{2004A&A...417..751A} found (O/H)$_{\odot}$ = (4.60 $\pm$ 0.50)$\times10^{-4}$ and \citet{2004astro.ph..7366M} determined (O/H)$_{\odot} \sim$ 4$\times10^{-4}$. This downward trend in the Solar oxygen abundance implies that a smaller amount of oxygen can be incorporated into dust. If we adopt the new solar oxygen abundance of \citet{2004A&A...417..751A} as representative of the Solar abundance of oxygen, then $\sim$25\% of the oxygen in the Local Bubble could be depleted onto dust grains. However, this should only be a lower limit to the amount of oxygen depleted in grains since stellar nucleosynthesis is supposed to increase the oxygen abundance from the protosolar value.

%  For instance, Allende Prieto, Lambert, & Asplund (2001) determined (O/H) = (4.90+/-?)e-4 by analyzing the forbidden [OI] $\lambda$6300 line with a time-dependent 3D hydrodynamical model of the solar photosphere, and by taking into account the Ni I blend.  Holweger (2001) analyzed [OI] and OI lines with LTE models, and by including corrections for NLTE and granulation effects, he recommended a photospheric oxygen abundance (O/H) = (5.44+/-?)e-4. 

% More recently Asplund et al.  (2004) used [OI], OI, OH vibration-rotation and OH pure rotation lines and a 3D hydrodynamical model of the solar atmosphere for determining the solar oxygen abundance.  They explored NLTE effects and derived a solar oxygen abundance of (O/H) = (4.60+/-0.50)e-4.  This downward trend in the solar oxygen abundance points to a smaller amount of oxygen that could be incorporated into the dust.  If we adopt the new solar oxygen abundance of Asplund et al.  (2004) as the ISM total (gas + dust) reference abundance, we estimate that the oxygen dust fraction in the Local Bubble is O/H = 1.15e-4.  This oxygen dust fraction agrees with the maximum amount of oxygen that can be tied up into silicate and oxide grains:  O/H = 1.80e-4 (Cardelli et al.  1996).  Our estimate implies that ~25% of the oxygen in the Local Bubble could be depleted onto dust grains.

In contrast to the D/O and D/H ratios which seem to have different values outside and inside of the Local Bubble \citep{2003ApJ...599..297H,2004ApJ...609..838W}, our results are consistent with a constant and single-valued O/H ratio, both inside and outside the Local Bubble, up to log $N$(H\,I) $\sim$ 21.

The composition of the Local Interstellar Cloud (LIC) has been studied using pickup ions observed inside the heliosphere \citep{2004AdSpR..34...53G}. These authors report atomic densities that yield O\,I/H\,I = 4.32$\times10^{-4}$ compared to our value of 3.45$\times10^{-4}$. \citet{2003ApJ...599..297H} have shown in a study of D I/O I ratios for 14 sightlines that the variation in the LB was small, consistent with the measurement uncertainties. Unless one assumes that O I and D\,I abundances are correlated, contrary to the usual assumptions of astration, both the O\,I and D\,I abundances do not vary by large amounts.  Thus, the heliospheric results are indicative of abundances in the LB.  The uncertainties in the densities at the termination shock and filtration factors reported by \citet{2004AdSpR..34...53G} imply that the uncertainty in the ratio above is 21\%.  The agreement within the uncertainties between the heliospheric value and the spectroscopic value, which were obtained by very different techniques, is reassuring and indicates that large systematic errors (such as incorrect oscillator strengths) are not significant.   

\subsection{Implications of (O/H)$_{\rm LB}$ on the Galactic Infall Scenario}

The similarity of the value of O/H in the Local Bubble reported here with the values measured at larger distances \citep{1998ApJ...493..222M,2003ApJ...591.1000A,2004ApJ...613.1037C} is striking and may provide a limit on the infall of low metallicity gas into this part of the Milky Way.  The quantity of gas above and below the Milky Way plane is likely to be much less in the vicinity of the LB compared to that in the LISM studied by \citet{2004ApJ...613.1037C}. The average densities, defined as the column density divided by distance, for the LB sightlines reported here ranges from 0.003 to 0.08 cm$^{-3}$ with a median value of 0.02 cm$^{-3}$ (mean value is 0.03 cm$^{-3}$). In comparison, the sightlines used by \citet{2004ApJ...613.1037C} to obtain their average value, for $d <$ 800 pc and $n_{\rm H} <$ 1.0 cm$^{-3}$, range from 0.09 to 0.91 cm$^{-3}$ with a median value of 0.26 cm$^{-3}$ (mean is 0.34 cm$^{-3}$).  Although the average density is not a direct measure of the gas in a column height above or below the plane, it is a qualitative indicator of the amount of gas in each region.  Thus, for similar infall rates, dilution of metals will be  larger in the LB compared to the LISM. Using the uncertainties in the mean in this work and that of \citet{1998ApJ...493..222M} as limits, the maximum possible difference between the LB and the LISM is 0.32$\times10^{-4}$, 9.3\% of the LB value of O\,I/H\,I. If the quantity of gas in the vicinity of the LB is a factor of 3 below that beyond the LB, then by using simple dilution arguments and assuming infalling material with properties similar to that studied by \citet{2004ApJS..150..387S} in Complex C with O\,I/H\,I, 0.17 that of solar \citep[(O/H)$_{\odot}$ = 4.6$\times10^{-4}$ from][]{2004A&A...417..751A}, we find the fraction of infalling gas within a mixing time to be no more than about 15\%.  This modest amount of infall would have a  small effect on the D I abundance in the LB.  Using the value of D I/H I = 2.2$\times10^{-5}$ reported for the Complex C material by \citet{2004ApJS..150..387S}, we find that if the present LB value is 1.56$\times10^{-5}$ \citep{2004ApJ...609..838W}, the D/H ratio would have been increased by infall from a value of 1.46$\times10^{-5}$, a negligible difference.  It is unlikely that such modest infalls can be used to explain the difference between the LB D/O and D/H values and the lower values found for more distant sightlines with typically $d >$ 500 pc and log $N$(H\,I) $>$ 20.5 \citep{2003ApJ...599..297H,2003ApJ...586.1094H,2004ApJ...609..838W}.  If we had compared our value with that of \citet{2004ApJ...613.1037C}, which has smaller uncertainties, the infall limit would be even smaller. Finally, we note that a discussion of infall without including mixing and chemical evolution is extremely simplistic and is included here only to indicate the magnitude of the effects implied by the small diference between the LB and the LISM values.

\subsection{Implications of (O/H)$_{\rm LB}$ on the Population of O-bearing Grains}

The similarity between the O/H values for the LB and LISM also allows us to speculate about the nature of the grain population in the two environments. The sightlines studied here, together with the ones from \citet{1998ApJ...493..222M} and the ones from \citet{2004ApJ...613.1037C} discussed above, probe different average line of sight densities, $n_{\rm H}$. As discussed above, the average densities for the Local Bubble sightlines presented here are all $\le$0.1 cm$^{-3}$, whereas the densities beyond the LB are higher. The constancy of the O/H ratio across this density range may indicate that the population of oxygen bearing grains is similar in the different environments. In contrast, many other elements vary with density \citep[see the review by][]{2004oee..symp..339J}. \citet{1986ApJ...301..355J} discussed how the iron depletion varies between sightlines with different average gas density, using a simple model proposed by \citet{1985ApJ...290L..21S}. For densities less than about 0.1 cm$^{-3}$, they pointed out that the sightline would only pass through clouds (or density fluctuations) corresponding to warm diffuse gas. With increasing density to a few tenths of atoms per cm$^3$, sightlines would also pass through cool diffuse clouds, and at even higher densities, cool dense clouds with much lower spatial frequencies, would be sampled. Fitting their data to a model, they found that Fe is a factor of four less depleted in warm diffuse clouds than in cool diffuse clouds, and suggested that the dust distribution was likely different in the two different diffuse environments.  Even though Fe/H varies in this $n_{\rm H}$ range, O/H does not. It is possible that this change in the Fe depletion is due to the presence of Fe in grain mantles that are easily created and destroyed in the interstellar medium.  It is also possible that the oxygen bearing grains and the grains responsible for the changes in the iron depletion are different populations. 

A related question may be why does O/H remain invariant beyond the LB, while D/O and D/H do not \citep[][and references therein]{2003ApJ...599..297H,2004ApJ...609..838W}. In a recent examination of the question whether dust can explain the measured variations in the D/H ratio, \citet{2004astro.ph.10310D} points out that outside dark clouds, the oxygen in grains is thought to be mostly in the form of Mg$_x$Fe$_{2-x}$SiO$_4$. On this basis and using the interstellar abundance of Si \citep{2004oee..symp..339J}, he estimated (O/H)$_{\rm gas}$ $\approx$ 0.76$\times$(O/H)$_{\rm total}$, a value similar to that found here and by others \citep{1998ApJ...493..222M,2002ApJS..140....3M}. If the O-bearing grains are as described above and are robust against destruction in the warm diffuse and cold diffuse media, what are the Fe-bearing grains, responsible for the variable depletion? These grains may be the aromatic carbonaceous grains \citet{2004astro.ph.10310D} suggests are responsible for depleting deuterium and hence the variation in D/O and D/H ratios \citep[][and references therein]{2004ApJ...609..838W}. \citet{1999ApJ...517..292W} have already suggested that $\sim$60\% of cosmic Fe could be associated with a population of  small carbonaceous grains.

\citet{2004astro.ph.10310D} suggests that if the scenario discussed above is correct, the destruction of these grains by shocks with $v_s >$ 100 \kms~will release both D and Fe into the interstellar medium. J. J. Linsky (2005, in preparation) shows that the D and Fe abundances appear to be correlated, and C. Oliveira (2005, in preparation) has shown that D/O and D/H decrease with increasing density, for $n_{\rm H} >$ 0.1 cm$^{-3}$. Thus it is possible that the reason why O does not vary, unlike Fe and possibly D, is the robust nature of the O-bearing grains, as described above.  If future studies strengthen the evidence for dust depletion of D, they will also strengthen the hypothesis of distinct grain populations.

\section{Summary}
\label{sum}

We have used $FUSE$~data in conjunction with STIS observations and previously published O\,I and H\,I column densities to determine the O/H ratio in the Local Bubble. We find the weighted average (O/H)$_{\rm LB}$ = (3.45 $\pm$ 0.19)$\times10^{-4}$ (1$\sigma$ in the mean) for a total of eleven sightlines. Our result is in agreement with previous determinations of this ratio along sightlines probing gas well outside the Local Bubble \citep{1998ApJ...493..222M,2004ApJ...613.1037C}. 

Based on the similarity of the two ratios above, our results suggest that the net effect of the uncertain $FUSE$~bandpass O\,I $f$-values is not significant in the determination of $N$(O\,I) for these LB sightlines.
 
%The O/H ratio derived in this work can be combined with D/O derived for the Local Bubble by \citet{2003ApJ...599..297H}. The inferred (D/H)$_{\rm LB}$ seems to be in disagreement with the measured (D/H)$_{\rm LB}$. However, this disagreement is not significant when one considers the square root of the weighted average variance to compare the measured and inferred values.

%Our results favor the suggestion of infall of metal-poor gas in the local ISM to explain the discrepancy between O/H for short and long low density sightlines \citep{2004astro.ph..6385C}.

Taking into account the latest determinations of the Solar abundance of oxygen \citep{2004A&A...417..751A,2004astro.ph..7366M}, our results indicate that $\sim$25\% of the oxygen in the Local Bubble is tied up in dust grains.

In conjunction with other studies \citep{1998ApJ...493..222M,2004ApJ...613.1037C} our work suggests limits on infall of low metallicity gas in this part of the galaxy.  It also suggests that the populations of O-bearing grains are similar in both warm and cold diffuse clouds.

\acknowledgments

This work is based on data obtained for the Guaranteed Time Team by the NASA-CNES-CSA \fuse~mission operated by The Johns Hopkins University. Financial support to U. S. participants has been provided in part by NASA contract NAS5-32985 to Johns Hopkins University. %Support for French participation in this study has been provided by CNES.
Based on observations made with the NASA/ESA Hubble Space Telescope, obtained from the Data Archive at the Space Telescope Science Institute, which is operated by the Association of Universities for Research in Astronomy, Inc. under NASA contract NAS 5-26555. %These observations are associated with proposal 7296.
The profile fitting procedure, Owens.f, used in this work was developed by M. Lemoine and the French \fuse~Team. We thank Nicolas Lehner for helpful discussions regarding $N$(O\,I).

%\appendix

%\section{Appendicial material}

\bibliography{ms}
\bibliographystyle{apj}
%\nocite{*}

\clearpage

\begin{deluxetable}{lccccc}
%\tablecolumns{4}
\tablewidth{0pc}
\tablecaption{Stellar Properties \label{star_properties}}
\tablehead{ 
\colhead{Star} & \colhead{T$_{\rm eff}$} & \colhead{log~$g$} &\colhead{$d$} &\colhead{$l$} &\colhead{$b$}\\
\colhead{}  &  \colhead{(K)}             &\colhead{(cm$^{-2}$)} &\colhead{(pc)} &\colhead{}   &\colhead{}}
\startdata
WD\,0050$-$332 &36320 $\pm$ 200 &7.87 $\pm$ 0.04 & 58 & 299.14  & $-$84.12 	 \\
WD\,0232$+$035 \tablenotemark{a} & 55,000--57,000 & 7.3--7.5 & 70 & 165.97 & $-$50.27  \\
WD\,1254$+$223  & 39600 $\pm$ 200 & 7.80 $\pm$ 0.03& 67 & 317.26  & $+$84.75 \\
WD\,2152$-$548 & 45800 $\pm$ 600 & 7.78 $\pm$ 0.07 & 128 & 339.73  & $-$48.06 \\
\enddata
\tablecomments{$T_{\rm eff}$, log~$g$ and $d$ are from \citet{1997ApJ...480..714V} unless noted otherwise. $l$ and $b$ are from the $SIMBAD$ database. All white dwarfs have spectral type DA.}
\tablenotetext{a}{$T_{\rm eff}$, log~$g$, and $d$ from \citet{2000ApJ...544..423V}.}
%\tablenotetext{b}{$T_{\rm eff}$ and log $g$ from \citet{1994MNRAS.267..653B}, $d$ from \citet{1998ApJ...500L..41V}.}
%\tablerefs{(1) \citet{1999ApJ...517..399N}; (2) \citet{1997ApJ...480..714V}; (3) \citet{1996A&A...314..217D}}
%\tablenotetext{b}{From Vennes et al. 1997}
%\tablenotetext{b}{$T_{\rm eff}$ and log $g$ from \citet{1997MNRAS.287..705M}, $d$ from \citet{1998ApJS..119..207H}.}
\end{deluxetable}
\clearpage
%$$$$$$$$$$$$$$$$$$$$$$$$$$$$$$$$$$$$$$$$$$$$$$$$$$$$$$$$$$$$$$$$$$$$$$$$$$$$$$ 
\begin{deluxetable}{lccccc}
%\tablecolumns{4}
\tablewidth{0pc}
\tablecaption{Log of \fuse~observations \label{fuse_obs}}
\tablehead{ 
\colhead{Star} & \colhead{Program ID} & \colhead{Aperture\tablenotemark{a}}& \colhead{Mode} &\colhead{Time (ks)} & \colhead{Date} } 
\startdata
WD\,0050$-$332 & M1010101 & L & TTAG & 16.4 & 2000 Jul 4 \\
	       & P2042001 & L & TTAG & 8.5  & 2000 Dec 11 \\
%
%WD\,0131$-$163 & P2041201 & L & T & 8.7 & 2000 Dec 10 & 2.0.5\\	
%
WD\,0232$+$035 & P1040501 & M & HIST & 2.3 & 2003 Dec 7 \\
	       & P1040503 & M & HIST & 16.1 & 2004 Jan 2 \\
	       & P1040504 & M & HIST & 12.0 & 2004 Jan 6 \\				
%	
%WD\,0455$-$282 & P1041101 & M & T & 19.7 & 2000 Feb 3 & 2.0.5 \\	
%	       & P1041102 & M & T & 10.1 & 2000 Feb 4 & 2.0.5 \\
%	       & P1041103 & M & T & 17.7 & 2000 Feb 7 & 2.0.5\\
%
%WD\,0501$-$289 & P2041601 & L & T & 13.9 & 2000 Dec 5 & 2.4.1 \\
%
%WD\,0549$+$158 & P2041701 & L & T & 13.9 & 2000 Nov 4 & 2.4.1 \\
%	       & M1010303 & L & T & 24.6 & 2001 Feb 17 & 2.4.1 \\
%
%WD\,0715$-$703 & P2042101 & M & T & 10.2 & 2002 Mar 11 & 2.4.1\\
%	       & M1050701 & L & T & 14.1 & 2003 Aug 15& 2.4.1\\
%
%WD\,1234$+$481 & P2040901 & L & T & 12.5 & 2000 Dec 27& 2.4.1 \\
%		& M1052401 & L & T & 6.2 & 2003 Jan 20 & 2.4.1 \\
%	       & M1052402 & L & T & 12.2 & 2003 Mar 18 & 2.4.1 \\	       
%
WD\,1254$+$223 & M1010401 & L & TTAG & 6.3 & 2000 Mar 6  \\
	       & M1010402 & L & TTAG & 12.2 & 2000 Apr 29  \\
	       & P2041801 & L & TTAG & 9.9 & 2001 Jan 28 \\
	       & M1010403 & L & TTAG & 8.1 & 2001 Feb 7  \\
%
%WD\,1631$+$781 & P1042901 & M & T & 22.1 & 2000 Jan 18 & 2.0.5\\
%	       & P1042902 & M & T & 30.2 & 2000 Jan 31 & 2.0.5\\
%		& M1052801 & L & T & 8.9 & 2002 Oct 28 & 2.2.2\\
%	       & M1052892 & L & T & 5.3 & 2003 Jan 12 & 2.2.2\\
%	       & M1052803 & L & T & 5.3 & 2003 Feb 27 & 2.2.2\\
%
%WD\,2004$-$605 & P2042201 & M & T & 10.2 & 2001 May 21 & 2.4.1\\
%	       & P2042202 & M & T & 35.3 & 2001 Aug 31 & 2.4.1\\
%	       & P2042203 & M & T & 37.0 & 2002 Apr 11 & 2.4.1\\
%
%WD\,2111$+$498 & M1010703 & L & T & 3.5 & 1999 Oct 11 & 2.4.1\\
%	       & M1010704 & L & H & 5.7 & 1999 Oct 11 & 2.4.1\\
%	       & M1010706 & L & H & 4.7 & 1999 Oct 13 & 2.4.1\\
%	       & P1043601 & L & T & 28.7 & 2000 Jun 20 & 2.4.1\\
%	       & M1053202 & L & T & 7.9 & 2002 Sep 4& 2.4.1\\
%	       & M1053201 & L & T & 4.4 & 2002 Oct 27& 2.4.1\\
%
WD\,2152$-$548 & P2042301 & M & TTAG & 15.5 & 2001 May 22 \\
	       & M1051501 & L & TTAG & 4.7 & 2002 Sep 24 \\ 

\enddata
\tablenotetext{a}{M and L stand for MDRS and LWRS, respectively.}
%\tablenotetext{b}{T and H stand for TTAG and HIST, respectively.}
%\tablerefs{(1) \citet{1999ApJ...517..399N}; (2) \citet{1997ApJ...480..714V}; (3) \citet{1996A&A...314..217D}}
%\tablenotetext{b}{From Vennes et al. 1997}
%\tablenotetext{c}{From Dreizler \& Werner 1996}
\end{deluxetable}
%\clearpage
%$$$$$$$$$$$$$$$$$$$$$$$$$$$$$$$$$$$$$$$$$$$$$$$$$$$$$$$$$$$$$$$$$$$$$$$$$$$$$$ 

\begin{deluxetable}{lccccc}
%\tablecolumns{4}
\tablewidth{0pc}
\tablecaption{Log of STIS observations \label{stis_obs}}
\tablehead{ 
\colhead{Star} & \colhead{Program ID} & \colhead{Grating}& \colhead{Exp. Time (ks)} & \colhead{Date} &\colhead{$\Delta\lambda$ (\AA)}} 
\startdata
WD\,0050$-$332 & O4G101010 & E140H  & 2.018 & 1999 Jan 14 & 1170--1372 \\
	       & O4G101020 & E140H  & 2.116 & 1999 Jan 14 & 1170--1372 \\
WD\,1254$+$223 & O67K03010 & E140H  & 1.851 & 2001 March 14 & 1140--1335 \\
	       & O67K03020 & E140H  & 2.632 & 2001 March 14 & 1140--1335 \\
	       & O67K03030 & E140H  & 2.632 & 2001 March 14 & 1140--1335 \\
	       & O67K03040 & E140H  & 1.950 & 2001 March 14 & 1140--1335 \\
	       & O67K03050 & E140H  & 2.632 & 2001 March 14 & 1140--1335 \\
	       & O67K04010 & E140H  & 1.851 & 2001 March 22 & 1140--1335 \\
WD\,2152$-$548 & O4G105010 & E140M  & 2.171 & 1999 May 23 & 1140--1735 \\
\enddata
\tablecomments{All the observations were performed through the 0.2$\mfarcs\times0.09\mfarcs$ aperture.}
%\tablenotetext{a}{Although these data were not used in this work we mention them here for completeness. The analysis of the STIS data is presented in \citet{2000ApJ...544..423V}.}
\end{deluxetable}
\clearpage

\begin{deluxetable}{lccccc}
%\tablecolumns{4}
\tablewidth{0pc}
\tablecaption{Equivalent Width Measurements (m\AA) \label{eqwidth}}
\tablehead{ 
\colhead{$\lambda$ (\AA)} &\colhead{$f$-value}& \colhead{WD\,0050$-$332} &\colhead{WD\,0232$+$035} & \colhead{WD\,1254$+$223}& \colhead{WD\,2152$-$548}}
\startdata
1302.1685 & 4.80$\times10^{-2}$ & 100.2 $\pm$ 1.9 & 115.4 $\pm$ 2.0 & 56.2 $\pm$ 3.0 & 128.0 $\pm$ 5.4\\
1039.2304 & 9.07$\times10^{-3}$ & 49.1 $\pm$ 1.3  & 48.2 $\pm$ 1.0  & 13.7 $\pm$ 0.9 & 63.1 $\pm$ 2.5 \\
976.4481  & 3.31$\times10^{-3}$ & 25.5 $\pm$ 2.4  & 25.4 $\pm$ 1.8  & \ldots    & 39.2 $\pm$ 6.6 \\
971.738\tablenotemark{a}  & 1.38$\times10^{-2}$ & 41.5 $\pm$ 5.4  & 48.2 $\pm$ 2.8  & \ldots           & 76.6 $\pm$ 11.2 \\
948.6855  & 6.31$\times10^{-3}$ & 38.5 $\pm$ 4.3  & 32.4 $\pm$ 2.5  & \ldots           & 61.5 $\pm$ 10.9 \\
936.6295  & 3.65$\times10^{-3}$ & 25.2 $\pm$ 3.9  & 24.5 $\pm$ 1.9  & \ldots           & 49.6 $\pm$ 6.6  \\
929.5168  & 2.29$\times10^{-3}$ & 23.9 $\pm$ 4.3  & 10.3 $\pm$ 1.5  & \ldots           & 40.6 $\pm$ 8.6 \\
924.950  & 1.54$\times10^{-3}$ & 19.3 $\pm$ 4.1  & \ldots            & \ldots           & \ldots           \\
\enddata
\tablecomments{Wavelengths and $f$-values are from \citet{2003ApJS..149..205M}.}
\tablenotetext{a}{Triplet structure used, see reference above.}
\end{deluxetable}
%\clearpage

\begin{deluxetable}{lcccc}
%\tablecolumns{4}
\tablewidth{0pc}
\tablecaption{O\,I Column Densities (Log) \label{methods}}
\tablehead{ 
\colhead{Method} & \colhead{WD\,0050$-$332} &\colhead{WD\,0232$+$035} & \colhead{WD\,1254$+$223}& \colhead{WD\,2152$-$548}} 
\startdata
AOD & 15.15 $^{+~0.09}_{-~0.12}$ (924)& 15.01 $\pm$ 0.03 (976) &14.22 $\pm$ 0.03 (1039) & 15.35 $^{+~0.08}_{-~0.10}$ (936)\\
COG & 14.98 $\pm$ 0.04 & 14.96 $^{+~0.04}_{-~0.03}$ & 14.27 $\pm$ 0.04 & 15.52 $^{+~0.07}_{-~0.06}$ \\
PF & 15.09 $\pm$ 0.09 & 14.94 $^{+~0.05}_{-~0.04}$ & \ldots & 15.30 $^{+~0.13}_{-~0.09}$\\
\tableline
Adopted & 15.07 $\pm$ 0.09  & 15.00 $\pm$ 0.06 & 14.25 $^{+~0.06}_{-~0.05}$ & 15.45 $\pm$ 0.10 \\
\enddata
\tablecomments{AOD, COG, and PF stand for apparent optical depth, curve of growth, and profile fitting, respectively. The transition used to determine $N$(O\,I) with AOD is indicated in ().}
\end{deluxetable}
%\clearpage

\begin{deluxetable}{lccccc}
%\tablecolumns{4}
\tablewidth{0pc}
\tablecaption{Column Densities and O/H Ratios \label{ratios}}
\tablehead{ 
\colhead{Star} & \colhead{Distance (pc)} &\colhead{log $N$(H\,I)} & \colhead{log $N$(O\,I)}& \colhead{O/H$\times10^4$} &\colhead{Ref(H\,I), Ref(O\,I)}} 
\startdata
WD\,0050$-$332 & 58 & 18.57 $^{+~0.03}_{-~0.04}$ & 15.07 $\pm$ 0.09 & 3.16 $^{+~0.76}_{-~0.65}$ & 1, 1 \\
WD\,0232$+$035 & 70 & 18.48 $^{+~0.04}_{-~0.05}$ & 15.00 $\pm$ 0.06 & 3.33 $^{+~0.59}_{-~0.56}$ & 2, 1 \\
WD\,1254$+$223 & 67 & 17.85 $\pm$ 0.05 & 14.25 $^{+~0.06}_{-~0.05}$ & 2.51 $^{+~0.48}_{-~0.39}$& 1, 1 \\
WD\,2152$-$548 & 128 & 18.84 $\pm$ 0.05 & 15.45 $\pm$ 0.10 & 4.07 $^{+~1.16}_{-~0.95}$ & 1, 1\\
\tableline
\tableline
WD\,0642$-$166 & 2.6 & 17.81 $\pm$ 0.11 & 14.29 $\pm$ 0.05 & 3.02 $^{+~0.95}_{-~0.75}$ & 3, 4 \\
HD\,34029 & 12.5 & 18.22 $\pm$ 0.08 & 15.02 $^{+~0.19}_{-~0.13}$ & 6.31 $^{+~3.69}_{-~1.95}$ &5, 6 \\
WD\,1634$-$573 & 37 & 18.85 $^{+~0.06}_{-~0.07}$ & 15.51 $\pm$ 0.03 & 4.57 $\pm$ 0.75 &7, 8 \\
WD\,1314$+$293 & 68 & 17.93 $\pm$ 0.03 & 14.49 $\pm$ 0.04 & 3.63 $^{+~0.44}_{-~0.40}$&9, 9 \\
WD\,0501$+$527 & 69 & 18.18 $\pm$ 0.09 & 14.86 $\pm$ 0.04 & 4.79 $^{+~1.19}_{-~0.99}$ & 10, 10\\
WD\,2309$+$105 & 79 & 19.11 $\pm$ 0.03 & 15.67 $^{+~0.04}_{-~0.03}$ & 3.63 $^{+~0.44}_{-~0.34}$& 11, 11 \\
HD\,116658 & 80 & 19.00 $\pm$ 0.10 & 15.58 $\pm$ 0.10 & 3.80 $^{+~1.39}_{-~1.11}$& 12, 12\\
WD\,1029$+$537\tablenotemark{a} & 132 & 18.62 $\pm$ 0.05 & $\geq$14.93  & $\geq$2.04 &13, 1\\
\enddata
\tablenotetext{a}{$N$(O\,I) from \citet{1999ApJ...517..841H} determined from the $\lambda$1302 transition is in disagreement with the lower limit determined from the weaker $\lambda$1039 transition observed by \fuse~which we report here.}
\tablerefs{(1) This work; (2) \citet{2000ApJ...544..423V}; (3) \citet{1999A&A...350..643H}; (4) \citet{2003ApJ...599..297H}; (5) \citet{2002ApJ...571L.169V}; (6) \citet{2002ApJ...581.1168W}; (7) \citet{1996aeu..conf..241N}; (8) \citet{2002ApJS..140...91W}; (9) \citet{2002ApJS..140...19K}; (10) \citet{2002ApJS..140...67L}; (11) \citet{2003ApJ...587..235O}; (12) \citet{1979ApJ...228..127Y}; (13) \citet{1999ApJ...517..841H}.}
\end{deluxetable}
%\clearpage

\end{document}